# DEALING WITH CURIOUS PLAYERS IN SECURE NETWORKS

Liam Wagner
Department of Mathematics and
St John's College,

The University of Queensland
Brisbane, Australia

## ABSTRACT

*In secure communications networks there are a great number of user behavioral problems, which need to be dealt with. Curious players pose a very real and serious threat to the integrity of such a network. By traversing a network a Curious player could uncover secret information, which that user has no need to know, by simply posing as a loyalty check. Loyalty checks are done simply to gauge the integrity of the network with respect to players who act in a malicious manner. We wish to propose a method, which can deal with Curious players trying to obtain "Need to Know" information using a combined Fault-tolerant, Cryptographic and Game Theoretic Approach.*

## INTRODUCTION

The concept of a curious player in a network was first used to describe a method of information extraction. We examine how curious players use the network structure to extract information from other players. This behavior doesn't necessarily present a clear danger to the integrity of the network and thus could easily go unnoticed.

The current literature uses game theory to model the behavior of networks of this type to try and find and eliminate errors. We will introduce the use of games and fault tolerance to build the structure for our own work.

We introduce via well-established methods within mechanism design, a protocol that will outplay this type of behavior and place well-defined rules on each player. By standardizing the way in which each player receives and deals with packets of information we can place restrictions on each players set of possible moves. These restrictions and game design allow us to present a cryptographic and game theoretic methodology to uncover and eliminate curious players within secure networks.

## BYZANTINE AGREEMENTS IN SECURE COMMUNICATION

The *Byzantine Generals Problem* (BGP) is the original and most fundamental ways to reach agreement in a distributed system [2]. The best way to conceptualise the BGP is to use the example of the Byzantine army poised for attack [4]. The Army is comprised of divisions each commanded by a general. Having sent out observers the general must decide on a course of action. This must be a collective decision based on all the available facts and played out by each division in unison.

However in some cases there may be a traitor. We should also note that in the model presented in [4], the location of the commanding general, or for that matter the traitors does not need to be taken into account.

Broadcasting guarantees the recipient of a message that everyone else has received the same message. This guarantee may no longer exist in a setting in which communication is peer-to-peer and some of the people within this network are traitors. In this type of setting a *Byzantine Agreement* offers the next best thing to a broadcast.

Byzantine Agreements (BA) are used widely as a method for fault tolerance in distributed systems. A critical example of their use is in bus systems, where fault tolerance maintains aircraft reliability (Rushby [12]).

The original literature of Lamport [4,5] and Pease [10] creates the possibility of using such an idea to maintain secure communication. The use of a more formalized version of the Byzantine Generals Problem is investigated in section 3 of Wagner [14], using the developments of Pease and Lamport [10,4].

The development of Byzantine Agreements in a secure communication environment in Linial [6], provides us with a wide-ranging insight into how BA's can be used to establish protocols for secure communication.

## CURIOUS PLAYERS

The main danger in not seeking out curiosity is the possibility of attack by some third party who has undermined the network by having been dealt information from the curious player.

A curious player is often defined as being someone who simply seeks out information that they may or may not have the right/need to know. What we must do now is make a more formal effort to describe this behaviour.

- Traverse the network to uncover and collect as much information that is available throughout the system.
- Players use the trust and loyalty of other players with access to alternate information.
- Store all messages seen throughout the duration of the player's presence within the network.
- Traitors collaborate to extract as much information as possible from the network using multiple attempts.

Although this definition of a curious player may seem a little vague, one must remember that this type of attack is completely subversive. This also leads to the proposition that curious players are passing this information to some attacker outside the network. Opponents, which rely on this type of information retrieval, are highly motivated and have inside knowledge of the structure and organization of the network being attacked.

In a conventional network the grand designer may be unaware of the amount of information flowing to all the other players. Although many organizations have the concept of a registry operating to keep track of who is reading what, there are still many avenues for discovery. Curious players use the trust placed in them by loyal players to discover information by other means.

Currently the methods used to uncover traitors, have been focused on malicious players, whose intent is to destroy the integrity of the network [6]. This paper tries to recognize the significance of players residing in the network, using its structure to undermine the capacity of the grand designer to limit information dispersal to the outside world.

The possibility of having players inside a network becoming curious is too great. The danger, which is demonstrated by the mere possibility of an outside agent finding out information from a curious player, should be enough for network designers to consider using an active method of deterrent.

What the current literature doesn't provide for are measures, which could actively deter curiosity [6]. The active search for members within the network is something, which has long been over due. The acceptance by the grand designer that this type of situation will happen is a more viable option than a simple deterrent protocol as outlined in [6].

Other methods in secure communication search for users who try to recombine access structures, which they may not be permitted to do. This is also a useful proposition but doesn't actively seek out curiosity amongst those with access but who lack the need to know.

Without questioning all players on what they have distributed across the network the grand designer is helpless. The active and continual search for the likely traitors in this type of network adds a great deal to the general integrity testing and vetting processes of high security environments.

## MECHANISM DESIGN

We wish to reverse engineer the way in which this type of game is played out. Implementation theory or mechanism design allows us to examine the inverse approach so as to fix a set of outcomes and look at the game form which yields the required set of outcomes as equilibria [9].

To institute a scheme to outwit this type of behavior we must place certain rules on each player. By standardizing the way in which each player receives and deals with packets of information we can place restrictions on each players set of possible moves.

- Each Player keeps 2 information sets
    - Information gathered or created.
    - Information transferred by other players.
    - These two sets are partioned to allow for both digital signatures which are applied to documents and full documents.
- Maintenance of two information sets, which are all, transferred in unison to the grand designer.
- Grand Designer searches each set of values to determine who has been curious and not disclosed their entire second information set.

In order to visualize the information flow in within the network as a loyalty check is performed we have included figure 1. This figure shows how information, which originates within the confidential level, is passed upwards through the other two clearances and also to the grand designer. In this case the grand designer will act as a registry and also communicate with the higher levels to assert the loyalty of each player.

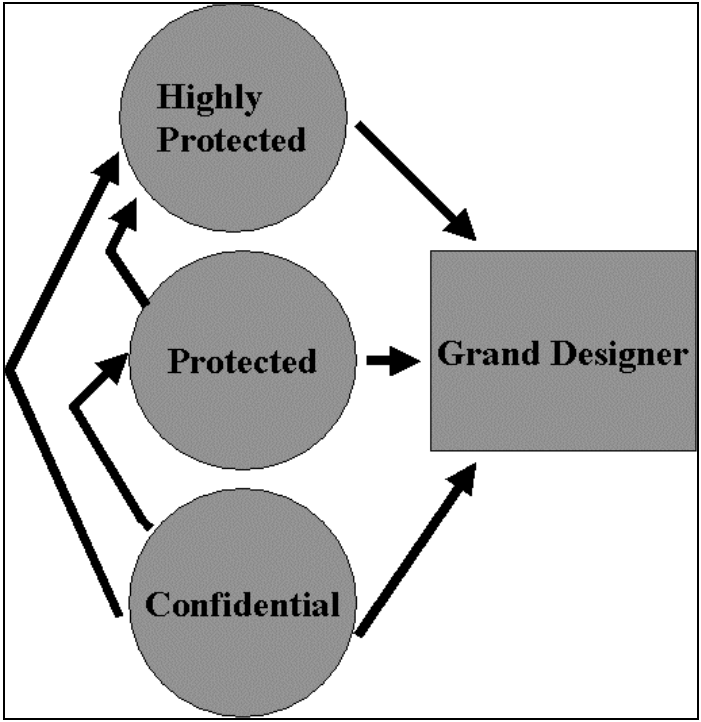

**Figure 1: Upward Information Flow**

## PROTOCOLS

The protocols, which we wish to implement in this situation, are rather simple and based around common practices within secure communication. We refer the reader to Menezes et al. [8] for a complete outline of the common methods used to encrypt information.

We shall now propose a procedure for the safe transmission of inter-clearance level information to be used in loyalty checks. This method blinds the information set of each player while still allowing for a direct comparison. Even though the information has been digitally signed and thus unreadable, one can still compare two information sets. If two information sets are identical then the digital signature, which is applied, should yield the same result. The combination of a digital signature and a document is unique and reproducible [8]. Thus one can still compare information sent around the network.

**Protocol for Inter-Clearance Levels**:
  I.   Bundle all Information (Catalogue Information).
  II.  Apply Digital Signature (Blinding Phase).
  III. Encrypt using a Public Key System.
  IV.  Transmission Phase (Sender)
       - Register information and pretext of transfer with grand designer.
  V.   Decrypt and Comparison Phase (Receiver)
       - Register receipt of information with grand designer.

We must formally introduce a procedure for members of a clearance level to check the information dispersal amongst their peers. This protocol allows for the full documents to be compared, which allows for people with the need to know a broader insight into the amount and type of information, which is being collected.

**Protocol for Intra-Clearance Levels:**
  I.   Bundle and Encrypt Information using Public Key System.
  II.  Transmission Phase (Sender)
       - Register information and pretext of transfer with grand designer.
  III. Decrypt and Comparison Phase (Receiver)
       - Register receipt of information with grand designer.

**Theorem 1.** The protocol reveals which members of the network are *curious players*.

**Proof.** Suppose that there exists a curious player. Since the transmission of any information is logged and registered, the curious players information set would be clear to other players. Furthermore each player must transmit his or her information set in a synchronous manner around the network. Thus for a curious player to succeed and evade the vetting of information dispersal they would have to convince other players of the inaccuracy of their peers registry. Due to the complex nature and size of a network, which obeys the Byzantine Agreement protocols in Wagner [14], a curious player would be out numbered by loyal players.

## RESULTS AND CONCLUSION

Systems like this have been implemented before for finding traitors in secure networks. But what make this work unique is that little has been done to engage the work of game theory to reveal traitors.

However the current literature, which explores secure networks, assumes that the network will be robust and reliable. This approach simply assumes that the traitors will be flushed out by other means [7]. These procedures don't allow for active search of curious players and the exclusion of traitors within the network via mechanism design.

These restrictions and game design will allow us to use a cryptographic and game theoretic solution to uncover and eliminate curious players within secure networks.